\documentclass[letter]{aa}

\usepackage{psfig}
\usepackage{graphicx}
\usepackage{natbib}
\usepackage[english]{babel}  
\usepackage{txfonts}

\begin{document}

\title{Spatial identification of the overionized plasma in W49B}

\author{M. Miceli\inst{1,2} \and F. Bocchino\inst{2} \and A. Decourchelle\inst{3} \and J. Ballet\inst{3} \and F. Reale\inst{1,2}}
\offprints{M. Miceli,\\ \email{miceli@astropa.unipa.it}}

\institute{Dipartimento di Scienze Fisiche ed Astronomiche, Sezione di 
Astronomia, Universit{\`a} di Palermo, Piazza del Parlamento 1, 90134 Palermo, 
Italy 
\and INAF - Osservatorio Astronomico di Palermo, Piazza del Parlamento 1, 
90134 Palermo, Italy 
\and DSM/DAPNIA/Service d'Astrophysique, AIM-UMR 7158, CEA Saclay, 91191 
Gif-sur-Yvette Cedex, France
}

\date{Received 23 November 2009, accepted 9 April 2010}

\authorrunning{M. Miceli et al.}
\titlerunning{Overionization in W49B}

\abstract{Recent $Suzaku$ X-ray observations of the ejecta-dominated supernova remnant W49B have shown that in the global spectrum there is a clear indication for the presence of overionized plasma whose physical origin is still under debate.}
{In order to ascertain the physical origin of such a rapidly cooling plasma, we focus on the study of its spatial localization within the X-ray emitting ejecta.}
{We confirm the presence of a saw-edged excess (interpreted as a strong radiative recombination continuum) in the global spectrum above 8 keV, emerging above the ionization-equilibrium model. We produce a hardness ratio map to determine where the plasma is overionized and we perform a spectral analysis of the regions with and without strong overionization.}
{We find that the overionized plasma is localized in the center of the remnant and in its western jet, while it is not detected in the bright eastern jet, where the expansion of the ejecta is hampered by their interaction with a dense interstellar cloud.}
{The location of overionized plasma suggests that the inner ejecta are rapidly cooling by expansion, unlike the outer ejecta, for which expansion is hampered by interstellar clouds seen in H$_2$}

\maketitle
\section{Introduction}
\label{Introduction}

X-ray observations of young supernova remnants (SNRs) allow us to probe the physical and chemical conditions of the shock-heated ejecta, and to study the physical processes involved in their interaction with the circumstellar ambient medium. 

W49B is one of the brightest ejecta-dominated SNRs observed in X-rays, where it shows a jet-like morphology. 
Its global X-ray spectrum, characterized by intense emission lines from He-like and H-like ions of overabundant metals (Si, S, Ar, Ca, Fe, and also Cr and Mn) indicates a plasma at collisional ionization equilibrium (CIE, see, for example, \citealt{fti95} and \citealt{hph00}). 
These results are confirmed by spatially resolved spectroscopy with \emph{XMM-Newton} (\citealt{mdb06} hereafter M06) which reveals a significant Ni overabundance ($Ni/Ni_{\odot}=10^{+2}_{-1}$) in the center. M06 found similar chemical composition and temperatures of the ejecta in the central and eastern regions, while in the western part the abundances (in particular for Fe) and the temperatures are lower. 
Both the comparison of the observed ejecta abundances with the yields predicted by explosive nucleosynthesis models (performed by M06) and the multipole expansion spatial analysis (performed by \citealt{lrb09}) concur in indicating a core-collapse origin for W49B.

The eastern border of the remnant shows bright radio emission (\citealt{llk01}), and is spatially coincident with a shocked molecular H$_{2}$ cloud that hampers the expansion of the ejecta in the east direction (being about three order of magnitude denser than the ejecta, see \citealt{krr07} and M06), thus distorting the eastern jet southward. Molecular hydrogen and [Fe II] and radio emission have been observed in the south-western edge of W49B, while in the center the [Fe II] emission reveals a barrel-like structure ``surrounding" the jet (\citealt{krr07}). \citet{mdb08} have shown that W49B seems to be the result of an aspherical jet-like supernova explosion with explosion energy $\la 1.5\times 10^{51}$ erg and mass of the shocked ejecta $\sim 6$ M$_{\odot}$. 

\citet{kon05} analyzed the global spectrum of W49B and measured the intensity ratio of the H-like to the He-like K$\alpha$ lines of Ar and Ca, finding that the ionization temperature $T_z$ is larger than the electron temperature, $T_e$, thus claiming the presence of overionization in W49B. 
M06 found $T_z>T_e$ for Ca and no evidence of overionization for Ar. Nevertheless, they also pointed out that the estimate of the electron temperature derived from the global spectrum is not reliable, given that it is significantly lower than that obtained from the spatially resolved spectral analysis of homogeneous regions. M06 also performed the same analysis on the uniform central region of W49B. Finding no evidence for overionization in the spectrum, they concluded that the temperatures of the Ar and Ca ions are consistent with $T_e$. 
Recently, \citet{oky09} (hereafter O09) analyzed the $Suzaku$ spectrum extracted from the whole remnant and revealed the presence of a saw-edged bump above 8 keV. They demonstrated that the bump is associated with a strong radiative recombination continuum of iron and produced an accurate spectral model to describe both the recombination continuum and the lines and to obtain diagnostics for the overionized ejecta (similar radiative recombination features have been also observed by \citealt{yok09} in IC 443, but for Si and S). O09 derived that, for Fe, the ionization temperature is $kT_z\sim2.7$ keV, while $kT_e$ is $\sim 1.5$ keV. Moreover, O09 estimated that the overionized plasma has the same emission measure as the bremsstrahlung emitting plasma, thus suggesting a common origin for the two components.
These results strongly indicate that the Fe ejecta in W49B are cooling so fast that they are overionized. Nevertheless, there are still a few important open issues.

First of all the electron temperature derived from the global spectrum by O09 is systematically lower than that observed in the uniform spectral regions discussed in M06, where $kT_e$ ranges between $\sim1.8$ keV and $\sim3$ keV, and in the accurately selected regions presented in \citet{lrp09} (based on $Chandra$ data), where $kT_e=1.8-3.7$ keV. This discrepancy in $kT_e$ may be due to the fact that the global spectrum originates from physically non-uniform regions of W49B with different temperatures and abundances.
Secondly, it is important to verify if all the ejecta in W49B are overionized. While M06 did not find overionization effects for Ar and Ca, O09 detected overionization for Fe. Notice that \citet{lrp09} have shown that the iron morphology is very distinct from that of other elements, being more asymmetric and more segregated (and localized in the central and eastern parts of the remnant), while Ar and Ca appear well mixed and more isotropic. So, in principle, it is possible that the physical conditions in the Fe-rich ejecta are different from those of the other ejecta.
Finally, a spatial localization of the overionized plasma may contribute to ascertain the physical origin of its rapid cooling, still not understood. 

Here we present the analysis of archive \emph{XMM-Newton} observations of W49B specifically devoted to constrain the spatial distribution of the overionized plasma. The data analysis procedure is shown in Sect. \ref{The Data} and our results are presented and discussed in Sect. \ref{Results} and Sect. \ref{Discussion}, respectively. 

\section{Data processing}
\label{The Data}

The \emph{XMM-Newton} data analyzed here consist of the two observations 0084100401 and 0084100501 (PI A.  Decourchelle) already presented in detail in M06 (see their Sect. 2 and Table 1). We re-processed the data using the Science Analysis System and following the same procedure and criteria described in M06. The images are a superposition of the pn images of the two observations and are background-subtracted, vignetting-corrected, and adaptively smoothed following the recipe described in M06. In the images, the background is subtracted by normalizing blank sky observations to the level of flux of the science observation in the 10-12 keV (12-14 keV) band for the MOS (PN) cameras.

For spectral analysis we used the same background event files used in M06 and the background spectrum was extracted from the same region positions on the CCD and subtracted from the source spectrum. The background flux is always a few percents of the signal in the $8-10$ keV band; as an example, for the pn camera in the spectral region Y (see Sect \ref{Results}), it is $\la4\%$ of the total signal in the $8-10$ keV band and $\la9\%$ of the total flux in the $8-12$ keV band. We first analyzed the spectra extracted from the two observations independently, and verified that they are compatible. Then, for each camera, we summed the spectra extracted from the two observations. Spectra were rebinned to achieve a signal-to-noise ratio per bin $>5\sigma$ and Ancillary Response Files were produced with the SAS $ARFGEN$ task. The spectral fits were performed simultaneously on the MOS1, MOS2, and pn spectra using XSPEC (\citealt{arn96}). 
We adopt the best-fit model described in M06. The model consists of two absorbed thermal component with oversolar abundances (Si, S, Ar, Ca, Fe, and Ni) in ionization equilibrium, plus two narrow gaussians to model the Cr and Mn lines. We are interested in the hard X-ray emission, so we fit the spectra in the $4.4-12$ keV band. M06 have found small scale variations in the best-fit parameters of the cold component and, since here we will focus on relatively large regions, we expect that these inhomogeneities may be an issue in the spectral fittings. However, in the hard $4.4-12$ keV band the cold component contribution is negligible. Since it is not possible to constrain the parameters of the cold component and the Si, S, Ar, and Ca abundances (whose emission lines are all below $4.4$ keV), we derive the best-fit temperature and emission measure of the cold component, the $N_{H}$, and the Si, S, Ar, and Ca abundances by fitting the spectra in the $1-9$ keV band and then we freeze them to their best-fit values when we perform the fits in the $4.4-12$ keV band. All the reported errors are at 90\% confidence.

\section{Results}
\label{Results}

\begin{figure}[htb!]
 \centerline{\hbox{     
     \psfig{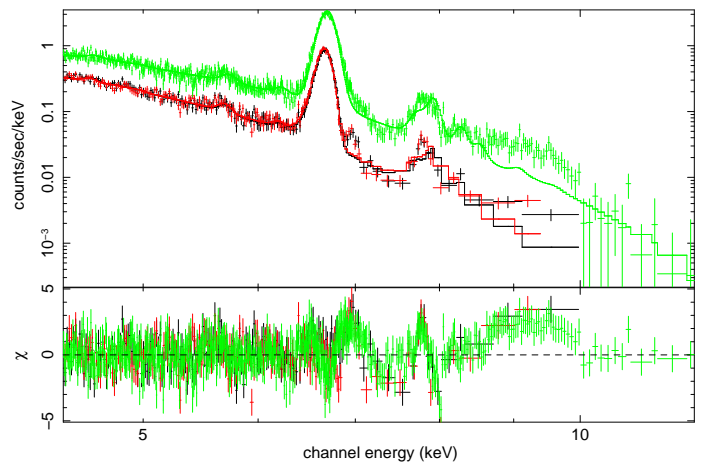}
     }}
\caption{pn (upper) and MOS (lower) global spectra of W49B in the $4.4-12$ keV band, together with their best-fit model and residuals. The best-fit model consists of two absorbed thermal components with oversolar abundances in collisional ionization equilibrium.}
\label{fig:spectot}
\end{figure}

We first extract the spectrum of the entire remnant and we model it in the $4.4-12$ keV band by adopting the best-fit model described in Sect. \ref{The Data}. We find a best-fit electron temperature $kT_2=1.72^{+0.03}_{-0.02}$ keV, quite similar to that reported by O09. 
The spectrum in Fig. \ref{fig:spectot} shows quite clearly the presence of a saw-edged excess above $\sim8.3$ keV, first associated with a radiative recombination continuum by O09. We also get a poor fit in the Fe (at $\sim 6.7$ keV) and Fe$+$Ni (around 8 keV) lines in qualitative agreement with O09. Even though the excess is weaker in the \emph{XMM-Newton} spectra than in the $Suzaku$ data (where there are several spectral bins at more than 5 sigmas from the model, see O09), still it is significant.
\begin{figure}[htb!]
 \centerline{\hbox{     
     \psfig{figure=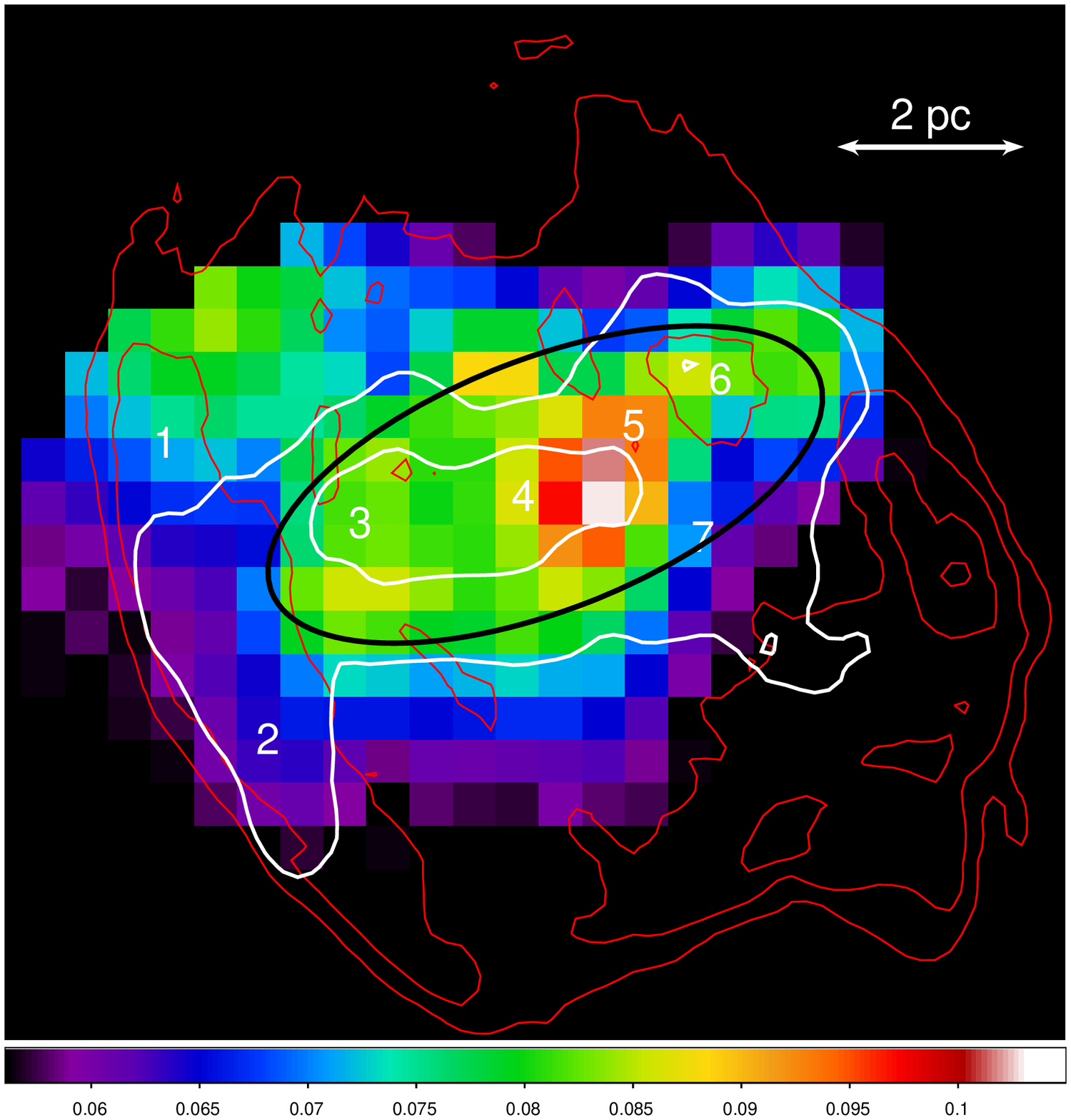,width=\columnwidth}
     }}
 \centerline{\hbox{     
     \psfig{figure=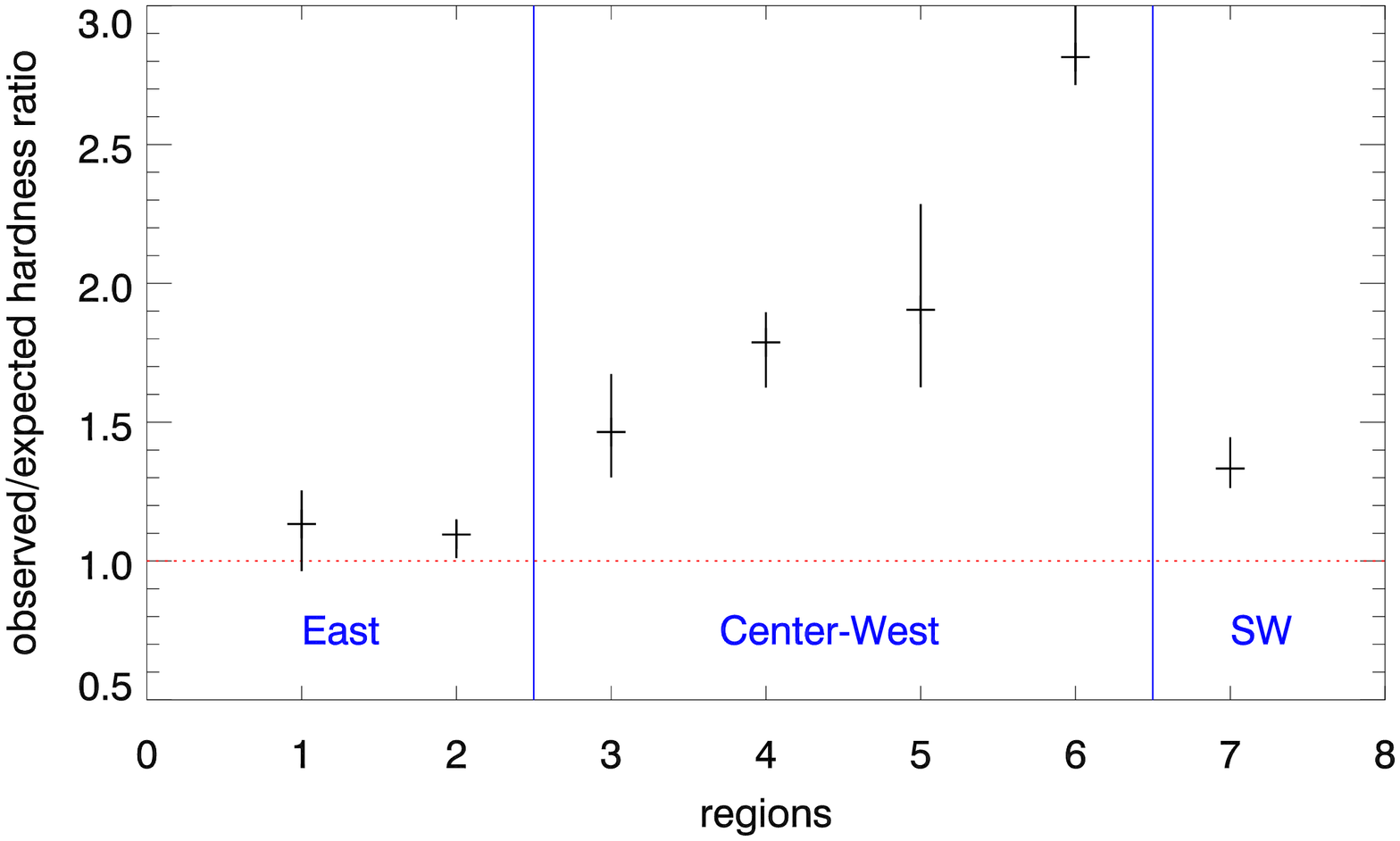,width=\columnwidth,angle=90}
     }}
\caption{\emph{Upper panel:} vignetting-corrected and background-subtracted pn map of the ratio of the count-rate in the hard ($8.3-12$ keV) and bremsstrahlung ($4.4-6.2$ keV) bands (i. e. H-S map). The image is adaptively smoothed to a signal-to-noise ratio 10 and the bin-size is $12''$. The white contour lines mark the EPIC count-rate (at $33\%$ and $66\%$ of the maximum) in the $1-9$ keV band, and the red contours are at $10\%$, $33\%$, and $66\%$ of the maximum of the radio image at 327 MHz (\citealt{llk01}). The region selected for the spectral analysis is bounded by the black ellipse. The 2 pc scale is obtained assuming a distance of 8 kpc. \emph{Lower panel:} ratio of the observed H-S values to that expected in the seven spectral regions (whose position is indicated in the upper panel) discussed in M06 assuming CIE.}
\label{fig:excess}
\end{figure}

To investigate the spatial distribution of the overionized ejecta, we produce a hardness ratio map, i. e. a map of the ratio of the pn count-rate in the high energy ($8.3-12$ keV) band to that in the bremsstrahlung ($4.4-6.2$ keV) band (upper panel of Fig. \ref{fig:excess}, H-S map). A high ratio indicates the presence of radiative recombination continuum and$/$or a hardening of the bremsstrahlung. We remark that if the overionization were not present we would expect large values of the H-S map in the eastern end of the remnant, where the ejecta temperature is larger (see M06 and \citealt{lrp09}). Instead, in our H-S map the low ratios are to the East. To quantify the deviations from CIE accounting for the temperature inhomogeneities, we calculated the H-S values expected assuming CIE in the seven regions discussed in M06, ($(H-S)_i^{CIE}$, $(i=1,...,7)$, computed using the $kT_{2,i}$ reported in Table 3 of M06) and compared them to those obtained in our H-S map, $(H-S)_i^{obs}$. The lower panel of Fig. \ref{fig:excess} shows the ratios $(H-S)_i^{obs}/(H-S)_i^{CIE}$. If the H-S variations were due to inhomogeneities in the plasma temperature, these ratios would be consistent with one, while values $>1$ indicate that the hard flux is larger than that expected in equilibrium of ionization. In regions 1 and 2 (at the eastern edge of the remnant) the hardness ratio is in agreement with that expected in CIE, while in regions 3-6 (at the central-western part of W49B) it is significantly larger. In region 7 (to the South-West) the difference is lower, but still significant. 
Figure \ref{fig:excess} therefore clearly shows that the hard radiative recombination continuum is stronger in the center and in the western region of W49B, while at the eastern edge the H-S values are consistent with CIE. As a cross-check, we also produced an ``excess image" (not shown here) by subtracting from the pn image in the $8.3-12$ keV band the underlying bremsstrahlung contribution (estimated by fitting the global spectrum in the $4.4-6.2$ keV band) and we obtained the same results.

The image analysis therefore indicates that the ejecta are overionized in the center and in the western region of W49B, while in the bright eastern edge of the remnant, where the ejecta are interacting with a dense molecular cloud, there is no significant excess. There are weak indications of overionization also in the south-western part of the remnant, characterized by strong radio and IR emission and faint X-rays. 

\begin{figure}[htb!]
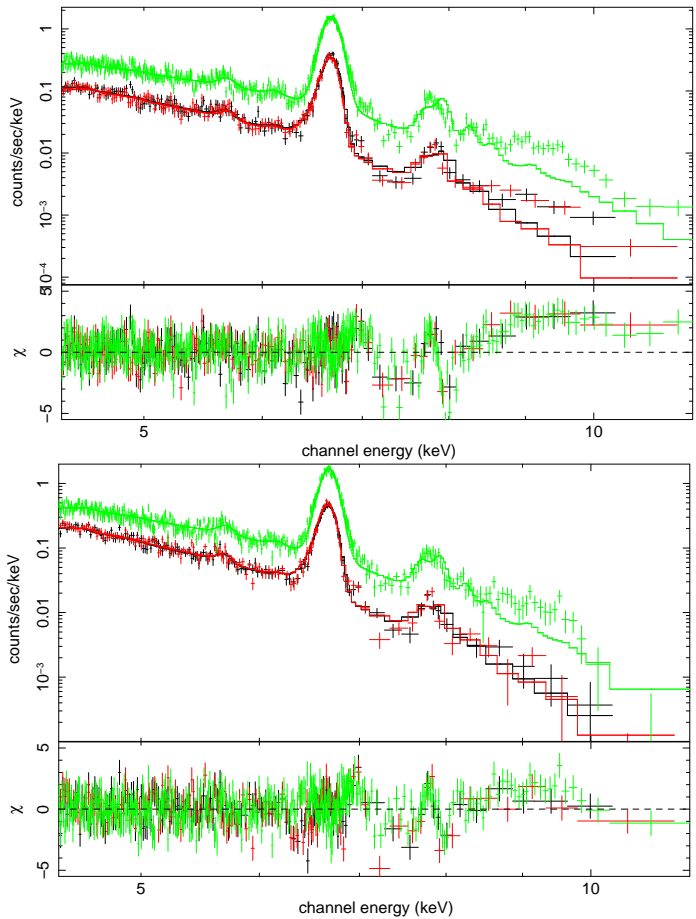

 \centerline{\hbox{     
     \psfig{figure=centrowest.ps,width=\columnwidth,angle=-90}}}
     \centerline{\hbox{ 
     \psfig{figure=tutto-cwreb.ps,width=\columnwidth,angle=-90}
     }}
\caption{\emph{Upper panel}: pn (green) and MOS (black and red) spectra of region Y (black ellipse in  Fig. \ref{fig:excess}) together with their best-fit model and residuals. \emph{Lower panel:} same as upper panel for region N (the complementary part of the remnant). We imposed the N spectrum to have the same spectral bins as the lower-statistics Y spectrum (where the signal-to-noise ratio per bin is $>5\sigma$). The best-fit parameters are shown in Table \ref{tab:specres}.}
\label{fig:spectra}
\end{figure}

As a further check we extract a spectrum from the black ellipse shown in Fig. \ref{fig:excess}, where we have localized strong overionization effects (we call this region ``Y'') and we compare it with the spectrum extracted from the complementary part of the remnant (i. e. the entire remnant $minus$ the overionized region, hereafter region ``N"). The two spectra are shown in Fig. \ref{fig:spectra} together with their best-fit models and residuals. To estimate the significance of the differences in the two spectra we imposed the N spectrum to have the same spectral bins as the lower-statistics Y spectrum (where the signal-to-noise ratio per bin is $>5\sigma$). The best-fit parameters are shown in Table \ref{tab:specres}. 

Figure \ref{fig:spectra} clearly confirms that the hard X-ray excess is dominant above $8.3$ keV in region Y, while it is much less significant in region N. The last row of Table \ref{tab:specres} also confirms that the excess is very large only in region Y. We remark that the low significance of the excess in region N is not due to statistics effects, since in this region the number of counts in the $4.4-12$ keV band is much larger than in region Y ($\sim 52000$ counts in region N vs. $\sim36000$ counts in region Y). 
Both the image and spectral analysis therefore indicate that the overionization effects are mainly localized in the central-western part of the remnant.

\begin{center}
\begin{table}[htb!]
\begin{center}
\caption{Best-fit parameters for the spectra extracted in region Y (black ellipse in  Fig. \ref{fig:excess}) and region N (i. e. the entire remnant minus region Y). The best-fit model and the fitting procedure are described in Sect. \ref{The Data}. All errors are at the 90\% confidence level.}
\begin{tabular}{lccc} 
\hline\hline
      Parameter                   &    Region   Y       &    Region N            \\ \hline
       $kT_{2}$ (keV)             & $1.73^{+0.03}_{-0.02}$ & $1.53\pm0.02$  \\
$EM^{*}_{2}$($10^{20}$ cm$^{-5}$) & $2.29^{+0.03}_{-0.6}$  &  $0.43^{+0.01}_{-0.01}$      \\
       $Fe/Fe_\odot$$^{(**)}$              & $5.8_{-0.2}^{+0.2}$ &      $4.7\pm 0.3$       \\    
       $Ni/Ni_\odot$$^{(**)}$              & $9.7_{-0.7}^{+1.5}$ &      $9\pm1$       \\           
reduced $\chi^{2}$$/$d.~o.~f.     &    $1.74/722$    &    $1.57/722$         \\   \hline
reduced $\chi^{2}/$d.~o.~f.($8.3-12$ keV)$^{(***)}$ &      $6.1/38$   &   $2.2/38$         \\ 
\hline\hline
\multicolumn{3}{l}{\footnotesize{* Emission measure per unit area.}} \\
\multicolumn{3}{l}{\footnotesize{** Abundances derived in CIE, Discrepancies with O09 are}} \\ 
\multicolumn{3}{l}{\footnotesize{due to the CIE model adopted here.}} \\
\multicolumn{3}{l}{\footnotesize{*** Contribution to the $\chi^2$ of the $8.3-12$ keV spectral bins.}} \\
\label{tab:specres}
\end{tabular}
\end{center}
\end{table}
\end{center}

\section{Discussion}
\label{Discussion}

Our analysis confirms the presence of a radiative recombination continuum in the high energy X-ray spectrum of W49B. As explained by O09, this indicates that the ejecta (at least the Fe ions) are overionized and are rapidly cooling to reach the ionization equilibrium. Nevertheless, we have shown that the bright eastern ejecta of W49B do not present tracers of overionization. It is important to stress that the eastern jet of ejecta cannot expand freely because a large and dense molecular cloud hampers the expansion in the east direction (\citealt{krr07}). The western jet, instead, is expanding undisturbed, since it does not interact with any dense structure. 

The overionization due to a rapid cooling of the ejecta is therefore observed only where the ejecta can expand freely. Such rapid cooling is then likely associated with the adiabatic expansion of the central-western jet. A similar scenario has been invoked by \citet{yok09} to explain the overionization effects in the supernova remnant IC 443.
We expect that the overionized Fe ejecta experienced a strong heating in the early phases of the SNR evolution. This is quite reasonable considering that the explosion occurred in a very complex environment, likely shaped by winds from the progenitor stars, and with the circumstellar material showing a barrel-like morphology (\citealt{krr07}). The interaction of the remnant with such environment can produce strong reverse and reflected shocks that can efficiently heat and ionize the ejecta. The subsequent rapid expansion of the ejecta in the center and to the West then produced their rapid cooling. We have estimated that a temperature $T\sim5\times10^7$ K and $\int n_edt\sim2.5\times10^{12}$ s cm$^{-3}$ would be necessary, in order to have equal ionization time (to H-like Fe) and electron heating time. Nevertheless, an accurate analysis requires a modeling of the hydrodynamic evolution of the plasma including the effects of non-equilibrium of ionization (e.g. \citealt{ro08}).

Not all the ejecta of W49B show signatures of overionization for the Fe ions and Fig. \ref{fig:excess} shows that in regions with similar (and large) Fe abundances, like the eastern jet and the central regions (see M06 and \citealt{lrp09}), there are different states of ionization (equilibrium to the East and overionization in the center), while the central and western regions (that have different Fe abundances, see M06 and \citealt{lrp09}) are both overionized. We conclude that, while we derive overionization effects for the Fe ions, not all the Fe-rich ejecta are overionized. On the other hand we remind that in the center of W49B no overionization effects are visible for the Ar and Ca ions (M06). 
We expect, therefore, different ionization temperatures and different ratios $kT_z/kT_e$ for different species and, at least in the case of Fe, for different regions of the remnant. In principle, it is also possible that the temperature of the electrons responsible for the Fe recombination is different from that of the electrons responsible for the bremsstrahlung emission in the $4.4-6.2$ keV band. However, from the global spectrum of W49B, O09 found that both the radiative recombination and the bremsstrahlung continua can be associated with electrons at the same temperature. Nevertheless, a spatially resolved study appears necessary in future works.

Another important issue is related to the determination of the value of the ion and electron temperatures in the ejecta. O09 have modeled both the radiative recombination continuum and line emission by deriving the ionization temperature from the global spectrum, finding $kT_z\sim2.6$ keV that is larger than the electron temperature $kT_e\sim1.5$ keV. In our large regions we also find similar values of $kT_e$, as shown in Table \ref{tab:specres}. Nevertheless, these values are significantly lower than those derived from smaller regions by M06. For example, our region Y is approximately the union of regions 3, 4, 5, and 6 in M06, where they find $kT_e\sim1.8-2.6$ keV that is significantly larger than the value $kT_e\sim1.75$ keV reported in Table \ref{tab:specres}. It is then important to analyze spectra extracted from relatively small regions in order to constrain accurately the ion and electron temperature distribution. Unfortunately, the available statistics does not allow us such a detailed analysis and we then stress the importance of obtaining deeper observations of W49B to address this point.





\bibliographystyle{aa}


\end{document}